\documentclass[11pt,a4paper]{article}
\makeatletter{} \usepackage{multirow}
\usepackage{microtype}
\usepackage{hhline}
\usepackage{xspace}
\usepackage{xcolor}
\usepackage{amsmath}
\usepackage{amsthm} 
\usepackage{euscript, stmaryrd, bigstrut}
\usepackage{wasysym}
\usepackage{algorithm}
\usepackage{algorithmic}
\usepackage{caption}
\usepackage{xfrac}
\usepackage{wrapfig}
\usepackage{bbm}

\usepackage{etoolbox} 

\usepackage[utf8]{inputenc}
\usepackage[english]{babel}
 
\usepackage[square,numbers]{natbib}
 \usepackage{amssymb}
 \usepackage{graphicx}  \usepackage{listings} \usepackage{mathrsfs}
 \DeclareMathAlphabet{\mathantt}{OT1}{antt}{li}{it}
 \DeclareMathAlphabet{\mathpzc}{OT1}{pzc}{m}{it}
 \usepackage[normalem]{ulem}  
 
 \usepackage{proof}
\usepackage{todonotes}

\usepackage{leftidx}
\usepackage{cmll}
\usepackage{subfig}
\usepackage{epigraph}
\usepackage{afterpage}
\usepackage{geometry}
\usepackage{textcomp}

\usepackage{float}
\floatstyle{boxed}
\restylefloat{figure}

\makeatletter{}

\newcommand{\qfbv}{the theory of quantifier-free bitvectors\xspace}

\newcommand{\mcl}[1]{} \newcommand{\mlc}[1]{}

  \newcommand{\wcl}{w-clause\xspace}
\newcommand{\wcls}{w-clauses\xspace}

\newcommand{\genExp}{\quad \overset{\varepsilon}{\rightsquigarrow}\quad}

\newcommand{\oneseq}{\mathbbm{1}}

\newcommand{\len}[1]{\|#1\|} 

\newcommand{\permof}[2]{\{#1\}=\{#2\}} 
\newcommand{\binary}[1]{\llbracket #1 \rrbracket}
\newcommand{\bneg}[1]{\overline {#1}} 
  \newcommand{\concat}[2]{#1 :: #2} \newcommand{\extract}[3]{\lfloor#1\rceil_{#2\ldotp#3}}

\newcommand{\zomit}[1]{}

\newcommand{\quadd}{\quad \quad}
\newcommand{\quaddd}{\quadd \quad}

\newcommand{\bv}{bitvector\xspace}
\newcommand{\bvs}{bitvectors\xspace}

\newcommand{\xor}{\oplus}
\newcommand{\band}{~ \& ~}

\newcommand{\ie}{\emph{i.e., \xspace}}

\newcommand{\eg}{\emph{e.g.,\xspace}\xspace}
\newcommand{\etc}{\emph{etc.}\xspace}

 \newcommand{\Escr}{{\cal E}}         \newcommand{\Mscr}{{\cal M}}

\newcommand{\Sscr}{{\cal S}}
  
\newcommand{\Cscr}{\text{$\mathcal{C}$}\xspace} \newcommand{\Lscr}{{\cal L}} 
\newcommand{\zc}[1]{\todo{Z}{\color[rgb]{1.000000,0.509804,0.278431} [ZC: #1]}}
\newcommand{\nsb}[1]{\todo[color=brown]{S}{\color[rgb]{0.000000,0.509804,
0.278431 } [SB: #1]}}

\makeatletter{}

\newtheorem{thm}{Theorem}[section]
\newtheorem{prop}{Proposition}[section]

\newtheorem{definition}{Definition}[section]
\AtEndEnvironment{definition}{$\bullet$}

\AtEndEnvironment{exmp}{$\odot$}

\theoremstyle{remark}

\AtEndEnvironment{rem}{$\therefore$ End of remark $\therefore$}

\renewcommand{\paragraph}[1]{\medskip\vspace{1mm}\noindent {\bf #1.}}

\renewcommand{\zc}[1]{}
\renewcommand{\todo}[2][\ ]{}    \renewcommand{\nsb}[1]{}

\title{CDCL-inspired Word-level Learning \\ for Bit-vector Constraint 
Solving\thanks{Work  partially funded by ANR 
under grants ANR-14-CE28-0020 and ANR-12-INSE-0002.}}

\author{Zakaria Chihani, Fran\c{c}ois Bobot, S\'ebastien Bardin\\
CEA, LIST, Software Security Lab, Gif-sur-Yvette, France\\(first.last@cea.fr)}

\date{}

\begin{document}
\maketitle


\makeatletter{}

\colorlet{lprolog}{blue!70!black}

\lstset{
  includerangemarker=false,
 numbers=left,
  firstnumber=0,
 language=[Objective]Caml
}

\begin{abstract}
\makeatletter{}

The theory of quantifier-free bitvectors is of paramount importance in 
software verification. 
The standard approach for  satisfiability checking reduces the bitvector problem 
to a Boolean problem, leveraging the powerful SAT solving techniques and their  
conflict-driven clause learning (CDCL) mechanisms.   Yet,  this bit-level 
approach  loses the structure of the initial bitvector problem. 
We propose a conflict-driven, word-level, combinable constraints learning for 
\qfbv.   
This work paves the way to truly  word-level decision procedures for bitvectors,  
taking full advantage of  word-level propagations recently designed 
in  CP and SMT communities.

\end{abstract}

\section{Introduction} \label{sec:intro}
\makeatletter{}

\paragraph{Context} 
Since the early 2000's, there is a significant trend in the research community 
toward reducing verification problems to the  satisfiability (SAT) problem of 
first-order logical formulas over well-chosen theories (e.g.~\bvs, arrays or 
floating-point arithmetic), leveraging the advances of  modern powerful  SAT and 
SMT solvers \cite{kroening08decision, Marques-Silva99grasp, Moskewicz01chaff, 
barrett09sat}.   
Besides  weakest-precondition calculi dating back to the 
1970's~\cite{dijkstra1976discipline}, most major recent verification approaches 
follow this idea \cite{Clarke04tool, Godefroid12test, Henzinger02lazy, 
McMillan06lazy}.

 In the context of verification, it is of paramount importance to efficiently 
and faithfully 
represent basic data types found in any programming language, such as arrays, 
floating points, integers and -- as is the focus of this paper -- \bvs (\ie 
fixed-sized arrays of bits equipped with standard low-level machine 
instructions~\cite{kroening08decision}).

\paragraph{The problem} The standard approach for  satisfiability checking 
reduces the bitvector problem to a Boolean problem, leveraging the powerful SAT 
solving techniques and their  conflict-driven clause learning (CDCL) mechanisms 
\cite{DBLP:conf/dac/MoskewiczMZZM01,biere2009conflict}.  
Yet,  this {\it bit-level approach}   loses the structure of the initial 
bit-vector problem,  disallowing any possible high-level simplification other 
than at preprocessing and, in some cases, yielding   scalability issues.  
Recent efforts in both the CP and SMT communities have demonstrated the 
potential benefits  of {\it word-level reasoning} over bitvectors 
\cite{bardin10tacas,michel12ppcp,chihani17cpaior,wang16cpaior,cotton2010natural,
de2013model} (where structural information ``is not blasted into 
bits''~\cite{bruttomesso2007lazy}), but  these works are  mostly limited to the 
propagation step  while the strength of modern SAT solvers rely on their 
learning mechanism \cite{biere2009conflict}. Actually, two recent works 
\cite{wang16cpaior,zeljic2016deciding} do combine word-level propagation and 
learning for bitvectors, yet the learning is either bit-level 
\cite{wang16cpaior} or assignment-dependent \cite{zeljic2016deciding}  in the 
vein of NoGood Learning \cite{katsirelos05aaai}.

\paragraph{Goal and challenge} Our goal is precisely to design a CDCL-like 
technique for \qfbv, such that it is truly 
word-level, and such that the learning stems from an interaction between the 
constraints involved in the conflict instead of a particular assignment to 
variables that lead to a conflict. This learning should be able to work 
together with  recently developed word-level propagation techniques for 
bitvectors.
Furthermore, in the context of verification 
where several theories can cooperate, it is also desirable that the 
technique can be combined with learning methods over other theories.

\paragraph{Proposal and contributions}
 We propose to lift bit-level learning mechanism to the word-level. 
This learning only makes sense if the decisions are allowed to be 
word-level, propagation is word-level, the conflict detection is word-level, 
the explanations are word-level and the sound rules of inference that derive 
knew knowledge from these explanations are also word-level. 
{\it Our very main  contribution} is  a CDCL-style, word-level learning over 
\qfbv,  enjoying all good properties from the SAT-based CDCL: the learnt clauses 
\textit{are part of the \qfbv theory}, they are \textit{asserting}, they are 
\nsb{tous ces points devraient être en theoremes, + un nom epxlicite (argument 
optionnel de thm)} 
\textit{minimal} (only involve propagations that lead to the 
conflict and variables that actively participated in the conflict), and of 
course  adding them is a \textit{sound extension} of the original problem (they 
are implied by some of the original constraints).  
More specifically, to achieve this goal, we design the following original ingredients: 
\begin{itemize}
\item \vspace{-2mm}
Word-level explanations for propagations of logical and structural \bv 
operations, through \bv constraints, associated with masks 
indicating the positions that they fix in the variables (Sec. 
\ref{sec:explanations}).

\item \vspace{-2mm}
A sound extension to word level of the Resolution rule, using \bv 
variables and constraints (Sec.~\ref{sec:resolution}).

\item \vspace{-2mm}
An explanation selection algorithm that relies on bit-wise 
operations to minimize the number of learnt constraints 
(Sec.~\ref{sec:analysis}).

\item \vspace{-2mm}
Finally, our learning mechanism, being based on resolution, can be 
\textit{combined} with resolution-based learning on other theories (notably 
arithmetic), allowing to  learn also from  simplifications  
(Sec.~\ref{sec:simplif}) used at propagation~\cite{chihani17cpaior}. 
\end{itemize}

\paragraph{Discussion} This work paves the way to truly  word-level decision 
procedures for bitvectors,  taking full advantage of  the word-level 
propagations recently designed in  both CP \cite{bardin10tacas, michel12ppcp, 
chihani17cpaior, wang16cpaior} and SMT communities \cite{cotton2010natural, 
de2013model, zeljic2016deciding}. The presentation of our technique is 
high-level. The many important   design questions behind an efficient 
implementation  (e.g., heuristics for clause deletion, using first versus last 
Unit Implication Point~\cite{Marques-Silva99grasp}, explanation recording 
techniques~\cite{feydy13trics} -- forward, backward or clausal, etc.)  are left 
as future work -- but our current method is fully compatible with all of them.

\section{Background }  \label{sec:background}
\makeatletter{} 

We briefly present the main characteristics of Conflict-Driven Clause Learning  (CDCL), and 
give a high-level presentation of \qfbv.

\subsection{Conflict-driven clause learning}
\zc{A short figure, if there is space, giving a summary algorithm for CDCL}
A constraint is a formula dependant on some variables to 
which one wishes to find, if any, an assignment to the variables that makes the 
formula true. A clause is a special kind of constraint in the form of a 
disjunction of Boolean variables. We use the word ``constraint'' and 
``clause'' interchangeably.
A conflict is a state of affairs (empty domain for a variable, falsehood of an 
original constraint) that makes the problem unsatisfiable. Most solving 
algorithms follow a sequence: 1- propagate, 2- if there is a conflict then 3- 
do 
``some Process C'', 4- else (once propagation is exhausted) make a decision 
(assign value to a variable, or restrict domain of a variable, or choose truth value of a 
constraint, etc.), then repeat. 
Originally, ``Process C'' was simply backtracking to the previous decision, but 
very early on, the problem of multi-occurring conflicts (or 
\textit{trashing}~\cite{mackworth1977consistency}) lead to non-chronological 
backtracking algorithms. One of the most famous of these algorithm -- 
the conflict-driven backjumping~\cite{prosser1993hybrid} -- directly 
inspired~\cite{bayardo1997using}  CDCL~\cite{DBLP:conf/dac/MoskewiczMZZM01, 
biere2009conflict}, which deduces from a conflict some new facts (here: 
Boolean clauses) 
allowing to circumvent future  similar conflicts, significantly short-cutting the 
search.

\paragraph{Learning techniques: requirements and properties}
Learning algorithms are usually built on the following key elements 
\cite{biere2009conflict}:
\begin{description}
 \item{\it Explanations}: to analyse the conflict, the solver needs to ``know'' 
the 
reasons for all the propagations that lead to that conflict. This implies that 
each propagation can generate some form of \textit{explanation}, which is 
implied by the original set of clauses -- either a theory lemma or some logical 
entailment ({\it explanation soundness}).  
\item{\it Selection}: once a conflict is detected, the learning algorithm must be able to 
select, from those explanations, the ones that actively participated in the 
conflict. 
\item{\it Sound extension}: to ensure that the learning algorithm only generates 
constraints (or clauses) that are implied by the original set of constraints, 
the relevant explanations that are selected are used to derive knew knowledge 
through {\it sound} rules of inference. In the case of CDCL, 
this rule is {\it Resolution}~\cite{robinson65jacm}. Thus adding a constraint 
that is a result of this inference is a \textit{sound extension}. 
\end{description}

Besides 
these three key elements 
a few other 
properties are also
 desirable. 
{\it (1)~Theory inclusion}: if the learnt constraint is part of the theory of 
the original set of constraints, then there is no need to introduce  extra 
machinery (any non-theory included constraint must be supplied with dedicated 
propagation, conflict detection, learning, etc.) in order to proceed with the 
solving. 
{\it (2) Asserting clause}: in order to shortcut the search, the added 
(learnt) constraint should force, immediately after (non-chronological) 
backtrack,  a propagation that prevents the same 
decisions as those which generated the conflict.
{\it (3)~Minimality}: the number of the constraints added through 
learning can be significant, thus it is preferable to learn through
information relevant to the conflict at hand, and nothing more, in order to 
minimize the size of the added knowledge. The selection 
algorithm should therefore discard what is not relevant to 
the conflict. However, if it is possible to avoid \textit{more} conflicts 
(create more shortcuts) with the \textit{same} additional space consumption 
(\eg the same constraint), that would be an advantage.
{\it (4) Compositionality}: problems involving many theories are common and 
very useful to verification, it is therefore desirable that the different 
learning components of each theory can 
cooperate 
with each another. 
The learning we introduce here can 
naturally be used in a wider Resolution-based learning context.

\subsection{Quantifier-free Bit-vector theory} 
\label{sec:background:bv}

\paragraph{Theory} A \bv is a vector of binary values, 0 or 1 (as such, the 
representation of an integer can be considered a \bv). The theory 
of quantifier-free \bvs surrounds that data type with logical operations 
(bitwise disjunction ($\mid$), conjunction ($\band$), negation ($\bneg .$), 
\etc), structural operations (shift ($\gg, \ll$), extraction 
($\extract{.}{i}{j}$ with $j<i<\textit{size}-1$ ), concatenation ($\concat$), 
\etc), and modular arithmetic. 
A \bv constraints is 
of the form $a \textit{ binop } b = c$, where \textit{binop} is a binary 
operator, and of the form $\textit{unop } a = b$, where \textit{unop} is a unary 
operator. 
These constraints can be considered atomically or appear in a clause (a 
logical disjunction). 
We will present our algorithm in terms of unit clauses which atoms are \bv 
constraints. Through compositionality with Boolean Resolution, however, it is 
possible to treat the wider clausal setting (Sec~ \ref{sec:mixing-theories}). 

\paragraph{Bitvector domains inside a decision procedure} During a decision procedure, a bit in a given \bv variable can have one of 4 states, depending on previous decisions and propagations:  
set (1), cleared (0), not yet known ($?$) and \textit{conflicting} ($\bot$).
A \textit{conflicting bit}  is a 
bit that was fixed (cleared or set) and that a propagation requires to be fixed 
to another value (resp. set or cleared). A \textit{conflicting variable} is a 
variable containing at least one conflicting bit. In this paper, a conflict is 
detected  when a conflicting bit is created, which is somewhat different from detecting conflict when a 
clause becomes unsatisfiable. 
Interestingly, these bitvector domains are shared by different word-level 
bitvector solvers from CP \cite{bardin10tacas, chihani17cpaior, wang16cpaior} 
and SMT \cite{de2013model, zeljic2016deciding}. 
From a practical point of view, they can be implemented either by a list 
\cite{bardin10tacas}, a \textit{run-length encoding} (a representation dependent 
on bit-alteration)~\cite{zeljic2016deciding} or (much more efficient) by a pair 
of integers, with a clever use of machine operations~\cite{michel12ppcp}.

\newcommand{\propzo}[3]{(#1)\rightharpoondown^#2 #3}

\zc{For motivating example, waiting for Aleks (mcBV author)}

\makeatletter{}
\section{CDCL-inspired Word-level learning}  \label{sec:cdcl}
Following the CDCL scheme (Section \ref{sec:background}), we 
first introduce  word-level explanations for word-level  propagations over bitvectors (Sec. 
\ref{sec:explanations}), 
we soundly extend resolution rules to \bvs (Sec. \ref{sec:resolution}), 
 then we show how we 
analyse the conflict, select the 
explanations and derive new constraints from 
them through our extended resolution (Sec. 
\ref{sec:analysis}). Finally, we extend the explanation language and illustrate 
the compositionality of learning through Resolution 
in order to encompass simplifications in the solver (Sec. \ref{sec:simplif}).

\subsection{Word-level explanations}
\label{sec:explanations}

We consider as a source for explanation the information propagated through the 
bitvector abstract domain (Sec.~\ref{sec:background}), as it is clear that 
word-level learning should be based on  word-level decisions and propagations. 
Moreover, bitvector domains are already used in existing word-level bitvectors 
solvers from CP \cite{bardin10tacas, chihani17cpaior, wang16cpaior} and SMT 
\cite{cotton2010natural, de2013model, zeljic2016deciding}.

The main idea here is to extend  Boolean Resolution (the combination rule of 
CDCL)  to \bv. The first step  is to find clause-like structures that are really 
\bv constraints,  yet still enjoying some of the properties of Boolean clauses. 
In the following, We reuse notations of bitvector domains  from Section 
\ref{sec:background}. 
We also use the usual notation $b_i$ to mean the $i^{th}$ bit of $b$ 
(recall that indices start at 0 and that the least significant bit is the 
rightmost one).

\begin{definition}
A \textit{word-level literal} $L$ (or simply \textit{literal} when clear from 
context), when $b$ is a \bv variable, is given through the following rules: 
\[
 L ::= T \mid  \bneg T \quaddd 
 T ::= A \mid  A \gg A \mid  A \ll A \quaddd 
 A ::= b \mid \bneg b \mid  \textit{constant}
\]
where $\gg, \ll$ are logical shift operators. 
\end{definition}
We say that a \bv variable $b$ \textit{is set (resp. cleared) 
through} a list of 
literals $\Lscr$ if its domain is set (resp. cleared) 
wherever (\ie on all positions where) the domains of \textit{all} literals of 
$\Lscr$ are set  (resp. cleared), and we write $\propzo{\Lscr}{1}{b}$ (resp. 
$\propzo{\Lscr}{0}{b}$). 
For example, from a constraint  $a \band b = c$, where $a , b, c$ 
respectively have the abstract domains 
$\binary{1??1},\binary{01??},\binary{??01}$, we 
forward propagate into $c$, through 
$\propzo{c}{0}{c}$, clearing $c_1$, and  
$\propzo{a,b}{1}{c}$ setting $c_0$, and 
backward propagate 
through $\propzo{c}{1}{a}$ and $\propzo{c}{1}{b}$ setting $a_2$ and $b_2$, 
$\propzo{c,\bneg{a}}{0}{b}$ clearing $b_3$.
The constraint $a\gg n = b$ backward propagates using $\propzo{\bneg{\bneg b 
\ll 
n}}{0}{a}$ and $\propzo{b\ll n}{1}{a}$. 

\begin{definition}
 A \textit{word-level clause}, or \wcl is of the form $L=\oneseq$ where $L$ is 
a bitwise disjunction of $l$-sized literals and $\oneseq$ is a 
sequence of $l$ ones (\ie the \bv that is set on every bit). By analogy with 
a Boolean clause, we write the constraint ($L=\oneseq$ ) simply as the 
disjunction of literals $L$. 
\end{definition}

A \wcl is called a ``propagation clause'' when it is used in the explanation of 
a propagation. 
When a propagation takes place in the domain of a variable $x$ 
through $\propzo{a^1,\cdots, a^n}{0}{x}$ (resp. $\propzo{b^1,\cdots, 
b^m}{1}{x}$), where $a^i$ and $b^i$ are (possibly shifted) literals, it can be 
explained through the \wcl $\bneg x \mid a^1 \mid \cdots \mid a^n$ (resp.  $x 
\mid \bneg{b^1} \mid \cdots \mid  \bneg{b^m}$). 
Table \ref{tab:prop-constraints} summarizes the \wcls for the four main  \bv 
constraints. 

\begin{table}[htbp]
\begin{center}
\caption{Propagation constraints for $\&,\mid, \xor$, and  
shifts, where $\{\lozenge, 
\blacklozenge\}=\{\ll,\gg\}$}
\label{tab:prop-constraints} \begin{tabular}{ | c | c | c | c | }
  \hline
  $x \band y = z$ & $x \mid y = z$ & $x \xor y = z$ & $x \lozenge y = z$ \\ 
  \hline
  $\bneg z \mid x$ &  $\bneg x \mid z$ & $ z \mid x \mid \bneg y  $ & $  \bneg 
z 
\mid x \lozenge y $\\
  $\bneg z \mid y $ & $\bneg y \mid z$  & $ z \mid \bneg x \mid y  $ &  $ z 
\mid 
\bneg{x \lozenge y} $ \\
 $z \mid \bneg x \mid \bneg y $&  $\bneg z \mid x \mid y$ & $ \bneg z \mid x 
\mid y $    &  $  x \mid \bneg{z \blacklozenge y} $ \\	
  & & $  \bneg z \mid \bneg x \mid \bneg y$& $  \bneg x \mid \bneg{\bneg z 
\blacklozenge y}   $ \\
  \hline
  \end{tabular}
\end{center}
\end{table}

\begin{prop}[Explanation soundness]
 The propagation clauses of Table \ref{tab:prop-constraints} are implied by 
their respective \bv constraint.
\end{prop}

Consider the (on the right) logical simplification using classical equivalencies 
and de 
Morgan rules of the bit-blasting 
encoding~\cite{kroening08decision} (on the left) of  the constraint $x \band y 
= z$, where $v_0,\cdots ,v_{l-1}$ are the  Boolean variables representing the 
$l$ long \bv 
variable $v$:
\[
\bigwedge_{i=0}^{l-1} ((x_i 
\wedge y_i) \longleftrightarrow z_i)
\quadd \equiv \quadd
\bigwedge_{i=0}^{l-1} ((\bneg{x_i} \vee 
\bneg{y_i}  \vee z_i) \wedge 
(\bneg{z_i}\vee 
x_i) \wedge (\bneg{z_i}\vee y_i))
\]
From this, it follows that each \bv conjunction constraint implies three other 
constraints that are really \wcls: \vspace{-2mm}
\[
 (x \band y = z) \quad\Longrightarrow \quad ( z \mid \bneg x \mid \bneg 
y=\oneseq)
                         \quad \wedge \quad (\bneg z \mid x=\oneseq) 
                        \quad \wedge \quad  (\bneg z \mid y=\oneseq) 
\]
A similar argument 
can be made for the other constraints of Table~\ref{tab:prop-constraints}. 

Note that \textit{the same} \wcl can serve as 
a propagation explanation for \textit{more than one} constraint. For example, 
$z \mid \bneg x \mid y$ 
explains setting $z$ wherever $x$ is set and $y$ is cleared from the 
constraints $z=x \band \bneg y$, 
$z=x \xor y$, $x=y\mid z$, \textit{etc}.

\paragraph{Explanation triples}
To be useful, each explanation must record two pieces of information : 
\textit{why} the propagation took place -- the propagation \wcl, call it \Cscr 
-- and \textit{when} it took place -- by indicating the \textit{decision 
level or depth} 
$p$ of the propagation (the number of decisions that preceded this 
propagation). 

In addition, because we deal with \bv variables of which some bits may be 
\textit{unaffected} by the propagation, we need a third piece of 
information $m$ we call \textit{impact mask} which is set for those 
bits of the variable that \textit{were unknown before} and that were cleared or 
set through the propagation.

\begin{definition}
 An \textit{explanation triple} $(p,m,\Cscr)$ explains the propagation at 
decision level $p$ on a variable along the positions set in the mask $m$  
using the propagation clause $\Cscr$. The set of explanations of a variable $x$ 
is written $\Escr(x)$. We say that an explanation \textit{covers} a bit (or a 
position) if its impact mask is set on that position.
\end{definition}

We emphasize the following remarks. Firstly, because the impact mask only 
indicates the bits that were unknown and became fixed,  
\textit{no two explanations can cover the same bit}: the order in 
which the propagations are done imposes this separation. 
For example, from the two constraints $e\band c = a, f \mid b = c$, with 
$a=\binary{1110}$, $b=\binary{0111}$ and $e,f,c$ are unknown, at a decision 
level $p$, both constraints can set the two middle bits of $c$, through the 
propagation \wcls $\bneg a | c$ or $\bneg b |c$ respectively. However, only one 
of them (the first to ``fire'') will be in an explanation that covers the two 
middle positions, \ie either $\{(p,\bneg a | c, 1110), (p, \bneg b |c, 
0001)\} \subseteq \Escr(c)$ or $\{(p,\bneg a | c, 
1000),(p, \bneg b |c, 0111)\} \subseteq \Escr(c)$.

Secondly, consider the constraint  $C: x \mid y = z$, where $x$ and $z$ are 
unknown, $y=\binary{00??}$. After a decision at level $p$ on some other 
variable, the 
propagations can follow this ordered sequence: $x$ gets cleared by some other 
constraint ($x=\binary{0000}$), which clears $z$ on some bits 
($z=\binary{00??}$) through $C$, then $y$ gets cleared by some other constraint 
($y=\binary{0000}$), and then $z$ gets completely cleared ($z=\binary{0000}$). 
Notice that the 
explanations for $z$ will both have the same propagation \wcl: 
$(p,1100,x\mid y \mid \bneg z),(p,0011,x\mid y \mid \bneg z)$. We choose 
to simplify such instances, \ie same propagation \wcl occurring \textit{at the 
same decision level}, to $(p,1111,x\mid y \mid \bneg 
z)$ by disjuncting their impact mask. Thus there is always a unique pair 
$(p,C)$ for each decide depth and 
propagation \wcl in the set of explanations of a given variable.

\subsection{Word-level resolution for  \wcls}
\label{sec:resolution}

A sequence of resolution steps involving disjoint set of clauses and variables 
can be carried out in parallel. Because a \wcl is essentially a 
group of Boolean clauses, they can subject to a word-level Resolution rule. 
Indeed, $l$ 
resolution steps between clauses involving the Boolean units $a_i,b_i,c_i$ (for 
$i=0..l-1$) can 
be considered as a word-level rule of Resolution between 
\wcls involving $l-$bit long variables $a, b $ and $c$.
\begin{center} \vspace{-6mm}
 \begin{align*}
\frac{a_0\vee c_0 \quadd \bneg{c_0} \vee b_0}{a_0\vee b_0} ,
\frac{a_1  \vee c_1 \quadd  \bneg{c_1} \vee b_1}{a_1 \vee b_1},
\cdots
,
\frac{a_{l-1}  \vee c_{l-1} \quadd \bneg{c_{l-1}}\vee 
b_{l-1}}{a_{l-1} \vee b_{l-1}}
\equiv 
\frac{a|c \quadd \bneg c|b}{a|b}
 \end{align*}
\end{center}

\smallskip

\begin{prop}[Soundness of inference]
 Word-level Resolution
 is a sound extension of  binary Resolution. 
\end{prop}

Word-level resolution on a shifted literal is less immediate. Consider that a 
shift constraint $ a \gg n = b$ is 
equiprovable with the Boolean encoding 
$(\bigwedge_{i=0}^{l-1-n} a_{n+i} \longleftrightarrow b_i) \wedge 
(\bigwedge_{i=l-n}^{l-1} 0 \longleftrightarrow b_i)$, which implies 
$(\bigwedge_{i=0}^{l-1-n} a_{n+i} \vee \bneg{b_i}) \wedge 
(\bigwedge_{i=l-n}^{l-1} \bneg{b_i})$, which in turn is 
equivalent and justifies the propagation \wcl $a \gg n \mid \bneg b$. The 
resolution can be extended to \bv constraints involving shift. One can, for 
example resolve $a \gg n \mid \bneg b$ with $\bneg a \mid c$, whose encoding is 
$(\bigwedge_{i=0}^{l-1} \bneg{a_i} \vee c_i$) to obtain $c \gg 
n \mid \bneg b$. Indeed, this \bv resolution can be simply seen as a 
group of $n$ Boolean resolution steps between $\bneg{a_{n+i}} \vee c_{n+i}$ and 
$a_{n+i} \vee \bneg{b_i}$	, for $i=0..l-1-n$, obtaining 
$(\bigwedge_{i=0}^{l-1-n} c_{n+i} \vee \bneg{b_i})$, which in conjunction with 
$(\bigwedge_{i=l-n}^{l-1} \bneg{b_i})$  is equivalent to $c \gg 
n \mid \bneg b$. More generally we can use the resolution rules:
\[
 \frac{A \mid (y \gg n) \quad B \mid \bneg y}{A \mid (B \gg n)} 
\quaddd \quaddd
\frac{A \mid (y \ll n) \quad B \mid \bneg y}{A \mid (B \ll n)}
\]

The word-level resolution rules acting on two shifted literals are justified 
in the same way. Of course, such a resolution only makes sens if the two shifts 
preserve some bits that are common to (that come from the same position of) the 
two literals, for example the literals $y\ll n$ and $\bneg y \gg m$, such that 
$n+m\geq \textit{size(y)}$ have no bits in common. 
Fortunately, the selection 
algorithm presented in the next section, by construction, only selects \wcls 
for which resolution is 
justified. The two literals can either have the same shift or opposing 
shift. The rules to apply depend on the length of each shift. In the following, 
$\{\lozenge, 
\blacklozenge\}=\{\ll,\gg\}$:
\[
 \frac{A\mid (y \lozenge n) \quad B \mid (\bneg y \lozenge m) \quad n > m}
         {A \mid (( B \blacklozenge m) \lozenge n} \quadd
 \frac{A\mid (y \lozenge n) \quad B \mid (\bneg y \lozenge m) \quad n \leq m}
         {B \mid (( A \blacklozenge n) \lozenge m} \quadd        
\]
Finally, resolution on two opposing shifted literals is as follows:
\[
 \frac{A \mid (y \gg n) \quad B \mid (\bneg y \ll m) \quad n > m}
{A \mid ( B \gg (m+n)) \mid (2^m-1) \ll (l-(n+m))}\quadd
 \frac{A \mid (y \gg n) \quad B \mid (\bneg y \ll m) \quad n \leq m}
{B \mid ( A \ll (m+n)) \mid (2^n-1) \ll m}
\]
notice that these rules subsume the above Resolution rules (if $n=0$ or 
$m=0$).

The logical and structural 
operations on \bvs all can be propagated using the 
basic blocks of conjunction, disjunction, negation and shift. 
For 
example, concatenation: $\concat a b =( a \ll \len b) \mid b$, extraction: 
$\extract{a}{i}{j}= (a\gg j) \band \bneg{(-1)\ll (i-j+1)}$, in addition to 
involutive negation of these operations: $ \bneg{\concat a b}=\concat{\bneg 
a}{\bneg b}$ and $ \bneg{\extract{a}{i}{j}}=\extract{\bneg a}{i}{j}$. 
If the literal's syntax is extended to include extraction and 
concatenations, one can also use the derivable rules of inference:
\[
 \infer{A \mid \extract{B}{i}{j}}{A \mid \extract{x}{i}{j}& \bneg x \mid B} 
\quaddd
 \infer{A \mid \concat B C}{A \mid \concat x C & \bneg x \mid B} \quaddd
 \infer{A \mid \concat C B}{A \mid \concat C x & \bneg x \mid B}
\]
However, for simplicity, we will present the conflict analysis and learning 
only through the basic blocks.

\paragraph{Normalizing} In order to obtain the actual \wcls, 
we normalize using (non-exhaustively listed) rewrite rules 
such as:
\begin{align*}
&\concat{(A \#  B)}{C} \longmapsto  
    (\concat A C) \#  (\concat B C)  
    &\quad&
    \extract{(A \#  B)}{i}{j}  \longmapsto 
    \extract{A}{i}{j} \#  \extract{B}{i}{j} 
    &\\	
    &\concat{C}{(A \#  B)} \longmapsto  
   (\concat C A) \#  (\concat C B)
   &\quad& 
   A \band B  \longmapsto A \wedge  B 
   &\\
   &(A \# B) \lozenge n \longmapsto A \lozenge n \# B \lozenge n 
   &\quad&
 \bneg{A \lozenge n} \longmapsto (2^n - 1) \blacklozenge (\len{A} - n) \mid 
 \bneg A \lozenge n &
\end{align*}
where $\# \in \{\band, \mid \}$ and $\{\lozenge, 
\blacklozenge\}=\{\ll,\gg\}$. 
\begin{prop}
 The normalization rewrite rules form a terminating rewrite system.
\end{prop}
Sketch of proof: By decreasing distance between the root and the logical connectives.

\paragraph{Mixing theories}
\label{sec:mixing-theories}
While this paper is concerned with purely \bv constraints, it is worth noting 
that the above \bv resolution rules can combine with usual resolution. 
\[
 \infer{
F \vee  (a \mid b) \vee G
 }{F \vee  (a\mid c) & (\bneg c \mid b) \vee G}
\]
In section \ref{sec:simplif}, we will use this fact to explain simplifications 
and factorizations, and include them in our analysis.

\begin{thm}[Learning soundness]
Any constraint obtained from the word-level resolution of \wcl (and Boolean 
resolution) is implied by the original problem and therefore constitutes a 
sound extension of it.
\end{thm}

\subsection{Word-level Conflict analysis}
\label{sec:analysis}

We recall that a conflict is detected when a conflicting variable has a 
conflicting bit (Section \ref{sec:background:bv}).
We now  show 
how word-level analysis is carried out, isolating the explanations for that 
conflict from the irrelevant explanations, resulting in conflict-driven 
word-level 
learning that is expressed inside the theory of quantifier-free \bvs, \ie 
through \bv constraints.

When the propagation of a constraint requires to set 
(resp. clear) one or more bits of a variable that were previously cleared 
(resp. set), a conflict is detected on those bits. A mask that is set on those 
conflicting positions is called a \textit{conflict mask}. For example, 
propagating from  $a \band  b = c$, with current domains 
$a=\binary{?1111111}$, 
$b=\binary{1?111111}$ and 
$c=\binary{00?00000}$ may generate a conflict through forward 
propagation resulting in 
$c=\binary{001\bot\bot\bot\bot\bot}$ or through 
backward propagation, resulting 
$a=\binary{011\bot\bot\bot\bot\bot}$ or 
$b=\binary{101\bot\bot\bot\bot\bot}$, in all these cases the conflict mask is 
00011111 and the aborted propagation clause is  $c \mid \bneg b \mid \bneg a$. 
This \wcl is called \emph{conflict \wcl}.

This \textit{grouped} detection exploits the word-level 
propagation as opposed to the conflict detection done through bit-blasting, one 
bit at a time. The main observation is the following: if \textit{several} 
conflicting bits are generated through\textit{ the same propagations from the 
same variables}, then there is no need to learn \textit{more than one} \bv 
constraint for those bits. In general more than one \wcl is learnt, possibly in 
parallel, after a 
single conflict.

\newcommand{\expeq}{=^\Escr}

\paragraph{Main idea}
For simplicity, we only discuss the algorithm for non-shifted literals (\ie 
variables) and we will then extend it.
The algorithm starts from the conflict at decide level $p$ caused by a 
propagation clause $R$ on the positions set in a conflict mask $m$ of some 
variable. 
The algorithm could be seen as applying Boolean CDCL in
parallel on all the set bits in the conflict mask $m$. 
At some points the
bits could have  different explanations and so the mask would be split
and the algorithm continues on two different inputs. 
Seeing each bit in 
isolation should be reminiscent of Boolean CDCL: each bit $b_i$ of each variable 
$b$ of $R$, such that the $i^{th}$ bit of $m$ is set, is necessarily fixed (a 
propagation through a \wcl on that bit has the same behaviour as a unit 
propagation through a Boolean clause: all other variables in the clause must be 
fixed).
We call such bits -- that participated in the conflict -- the \textit{culprit 
bits} of $b$. 
The values of the culprit bits were either fixed before $p$ (pre-$p$) or through 
a propagation at $p$ ($@p$). 
In the former case, there are no $@p$ explanations that cover $b_i$ ; in the 
latter case, there is exactly one explanation that covers it. 
We define the \textit{explanation equivalence} ($\expeq$) of bits 
$b_i$ and $b_j$ such that $b_i \expeq b_j$ if they are either both pre-$p$ or 
are both covered by the same $@p$ explanation in $\Escr(b)$. This 
equivalence relation is the one that fixes the number of eventually learnt 
clauses: there is a new clause each time there is more than one equivalence 
class on the bits of a variable.

\zc{Dessin avec des bits track\'es et \`a un moment une s\'eparation entre les 
explications}

\zc{Just a resolution with masks}

\begin{definition}
 A \emph{reason} $R$ is a \wcl that serves as an accumulator and is 
initialized 
with the conflict 
\wcl. Each reason is associated with an \emph{analysed mask} $m$, initialized 
by the conflict mask. For each equivalence class $E^=$ of the culprit bits 
of each variable in $R$, a separate copy of $R$ is created and associated 
with a new analysed mask which is set on the position of the bits in $E^=$. 
Once a variable has been investigated in this manner, it is 
\emph{marked}, such that it is purged from future resolution steps of the 
algorithm. We call a \emph{leaf} a reason whose literals are all marked 
(thus it is fully investigated). At the end of the analysis, all leaves are 
learnt. Finally, we call \emph{set of reasons} the set of tuples $(p,R,m^a, 
\Mscr)$ 
containing the reason $R$, associated analysed mask $m^a$,
decide depth $p$, with marked literals $\Mscr$. 
\end{definition}

\begin{algorithm}[htbp] \footnotesize \centering
 \begin{algorithmic}[1]
  \IF {$\exists x \in R, x\notin \Mscr$} \label{line:select}
   \IF {$\exists (p, C, mi) \in \Escr(x)$ s.t. $mi \band m^a \neq 
0$}\label{line:expWithIntersection}
     \IF {$mi \band m^a = m^a$} \label{line:allCovered} 
       \RETURN $\{(p,\textit{Resol(R,C,x,$\Mscr$)},m^a, \{x\}\uplus \Mscr)\}$ 
\label{line:allCovered-result}
 
\ELSE
       \STATE $m^{in}=mi \band m^a$  \label{line:min}
       \STATE $m^{out} =m^a \xor m^{in}$ \label{line:mout}
       \STATE $S=\textit{Resol(R,C,x,$\Mscr$)}$
       \RETURN $\{(p,S,m^{in}, \{x\}\uplus \Mscr),
             (p,R,m^{out}, \Mscr)\}$ 
    \ENDIF
   \ELSE \label{line:noExp}
       \RETURN $\{(p,R,m^a, \{x\}\uplus \Mscr)\}$ 
   \ENDIF
    \ELSE 
     \RETURN $\{\textit{leaf($R$)}\}$ 
  \ENDIF
\caption{$\textit{analyse}(p,R,m^a, \Mscr)$ (reason $R$, analysed mask $m^a$,
decide depth $p$, with marked literals $\Mscr$)}
 \label{alg:analyse}  \end{algorithmic} 
\end{algorithm} 
  
 \paragraph{Conflict analysis algorithm}
 After a conflict, the set of reasons starts as a singleton containing the 
tuple formed with the conflict 
propagation $R$, the conflict mask $m$, the current decision level  $p$ and an 
empty set that 
will hold the marked literals, \ie  $\{(p,R,m, \{\})\}$. The 
(parallelizable) algorithm consists of the repeated rewriting of the set of 
analysed reasons until all its elements become leaves, at which point they are 
all learnt. 
\[
 \{(p,R,m^a, \Mscr)\}\uplus \Sscr \longmapsto \textit{analyse}(p,R,m^a, 
\Mscr)\uplus \Sscr
\]

For now, we voluntarily omit shifted literals in Algorithm 
\ref{alg:analyse}, (we will see shortly how to easily extend it for shifted 
literals). We comment this algorithm through the following example, 
where, at decide level ($i$-1), values are $c=??11$, 
$v=1111$, 
$e=0000$, $g=00??$, $h=??00$, and unknown for the rest.
\begin{align*}
  &(1.) d=\bneg c\mid x& \quad  & (2.)e=b\xor f& \quad   &(3.)f = \bneg d \band 
v& \\
  &(4.)g=f\band a &  & (5.)h=f\band a & &  (6.)a=b\band c &
\end{align*}
Then, at level $i$, we decide $d=0000$. The following table lists the 
explanations 
stemming from that decision. We omit the decide level from the triples (all 
explanations are of the current decision level $i$) and replace it, for 
readability, 
with the index of the  constraint that spawned the propagation they explain.

\begin{center}\footnotesize
\begin{tabular}{ | c | c | c | c | }
  \hline
  $f$ & $b$ & $c$ & $a$ \\ 
  \hline
  (3, $f\mid d \mid \bneg v$, 1111) & (2, $b\mid e \mid \bneg f$, 1111)  &
   (1, $c\mid d$, 1100) & (4, $g\mid \bneg f \mid \bneg a$, 1100)  \\
   &&& (5, $h\mid \bneg f \mid \bneg a$, 0011) \\
  \hline
  \end{tabular}
\end{center}
At this point the $6^{th}$ constraint tries to propagate with \wcl $a\mid \bneg 
b \mid \bneg c$ and provokes a conflict on all bits. Conflict analysis then 
begins with the initial set of analysed reasons containing the tuple 
$\{(i,a\mid 
\bneg b \mid \bneg c,1111, \{\})\}$. 

\begin{figure}[htbp] \footnotesize
\centering
  \begin{tikzpicture}[node distance=1cm]
       > = stealth,             shorten > = 1pt,             auto,
            node distance = 1cm,             semithick         ]        
\node (a) {$a\mid \bneg b \mid \bneg c$ , 1111};
\node (br) [below right of=a, yshift = -0.5cm, xshift = 2cm]{$h\mid \bneg f 
\mid 
 \bneg b \mid \bneg c$, 0011};
\draw [->] (a) -- (br) node [midway, right] {a} ;
\node (bbr) [below of=br]{$h\mid \bneg f 
\mid  e \mid \bneg c$, 0011};
\draw [->] (br) -- (bbr) node [midway, right] {b} ;
\node (bbbr) [below of=bbr]{$h\mid d \mid \bneg v 
\mid  e \mid \bneg c$, 0011};
\draw [->] (bbr) -- (bbbr) node [midway, right] {f} ;

\node (bl) [below left of=a, yshift = -0.5cm,xshift = -2cm]{$g\mid \bneg f \mid 
\bneg b \mid \bneg c$, 1100};
\draw [->] (a) -- (bl) node [midway, left] {a} ;

\node (bbl) [below of=bl]{$g\mid \bneg f \mid 
e \mid \bneg c$, 1100};
\draw [->] (bl) -- (bbl) node [midway, right] {b} ;

\node (bbbl) [below of=bbl]{$g\mid d \mid \bneg v \mid 
e \mid \bneg c$, 1100};
\draw [->] (bbl) -- (bbbl) node [midway, right] {f} ;
\node (bbbbl) [below of=bbbl]{$g\mid d \mid \bneg v \mid 
e$, 1100};
\draw [->] (bbbl) -- (bbbbl) node [midway, right] {c} ;
\end{tikzpicture}
   \caption{Conflict analysis through rewriting and resolution}
 \label{fig:exmp}
\end{figure}
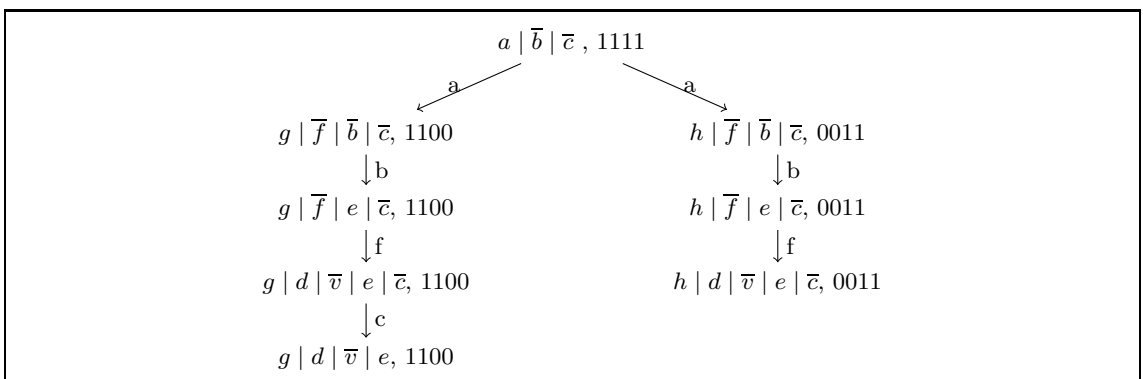

\paragraph{Unfolding the algorithm}
The algorithm selects the literal $a$ (Line \ref{line:select}), and the first 
selected explanation for $a$ is  $(i, g\mid \bneg f \mid \bneg a, 1100)$ 
since $1100 \band 1111 \neq 0$. The test of Line~\ref{line:allCovered} is not 
satisfied however because the impact mask of the propagation clause does not 
cover all the conflict mask. In the \textit{else} branch, two masks are 
created: one, $m^{in}$, will be the analysed mask of the reason resulting from 
the resolution of the propagation clause and the current reason, the 
other, $m^{out}$, represent the conflict bits that must be explained elsewhere. 
The algorithm returns two tuples: 
$\{
(i, g\mid \bneg f \mid \bneg b \mid \bneg c , 1100, \{a\}), 
(i,a\mid \bneg b \mid \bneg c,0011, \{\})
\}$. The former will be rewritten along the left branch of 
Figure~\ref{fig:exmp}, the latter -- where $a$ is \textit{not marked} -- 
will 
select $a$ again and this time chose the all covering explanation that comes 
from  the $5^{th}$ constraint. Once there are no more variables to select from a 
reason (\ie all variables 
are marked, Line \ref{line:noExp}), that reason 
is marked as a leaf. Interestingly, because the learnt constraint has a clausal 
form, it serves as its own propagation clause.

Figure \ref{fig:exmp} summarizes the learning through conflict 
analysis. The nodes 
are the reasons and the analysed mask (with the root being the initial pair). 
Each edge labelled with a variable name represents a resolution step on that 
variable between the parent reason and the propagation clause of the 
explanation of that variable that covers the resulting analysed mask. 

Notice how $\bneg c$ stays only in one of the learnt \wcls, because the right 
analysed mask is not covered by any 
current decision-level explanation of $c$, meaning that \textit{the part} of 
$c$ 
that 
\textit{participated }in the conflict \textit{predates }the decision, and 
indeed, initial value for $c$ was $??11$. 
The two learnt clauses propagate each 2 bits into $d$, avoiding the conflict.

This learning through interaction of constraints, independent  of a particular 
assignment, has the advantageous side effect of preventing word-level 
conflicts from potentially bit-level conflicts, \ie even a conflict mask with a 
single set bit will result in a learnt \wcl that impacts all positions of its 
variables, thus avoiding \textit{more than that exact} conflict with the 
\textit{same 
amount of added} space.

\paragraph{Different strategies}
The order in which the analysed reasons are rewritten by
\textit{analyse} can be tuned to obtain certain known strategies such as 
First-UIP or Last-UIP~\cite{Marques-Silva99grasp}.
The determined strategy can be implemented by the order induced by a trail.

\paragraph{Treating structural operations}
A slight addition to the above algorithm allows to treat shift operations. If 
the 
selected literal appears shifted by $n$, then the search for an explanation 
among the literal (line 3) uses the same shift of the impact mask. Line 
\ref{line:select}'s 
test becomes $\exists a\gg n \in R$, and line \ref{line:expWithIntersection}'s 
test 
becomes ``s.t $(mi \gg n) \band m^a  \neq 0$''. Other 
structural operations 
are also adapted accordingly, or supported through the rewriting that uses 
shift 
operations and logical operations.

\begin{figure}[htbp]
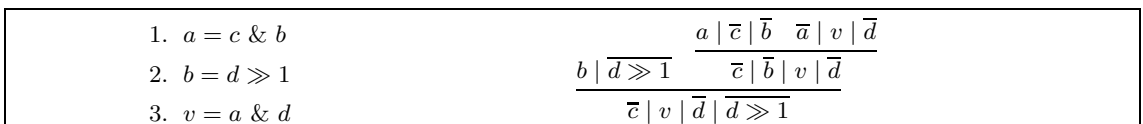
 \footnotesize
\centering
\begin{minipage}{0.41\textwidth}
 \begin{enumerate}
  \item $a = c \band b $
  \item $b = d \gg 1$
  \item $ v = a \band d$
 \end{enumerate}
\end{minipage}
\begin{minipage}{0.41\textwidth}
$
 \infer{\bneg c \mid v \mid \bneg d \mid \bneg{d \gg 1}}
 {b \mid \bneg{d \gg 1} & 
  \infer{\bneg c \mid \bneg b \mid v \mid \bneg d}
   {a \mid \bneg c \mid \bneg b & \bneg a \mid v \mid \bneg d}}
$
\end{minipage}
 \caption{Non-asserting clauses justify partial decision on variables}
 \label{fig:nonAsserting}
\end{figure}

\paragraph{Taming non-asserting learnt clauses}
To be useful, a  learnt clause must be \textit{asserting}, \ie force a 
propagation after a backtrack (or a back-jump) so as to avoid  the same 
conflict. With our method, this is not always the case. Consider the
example on 2-bits variables given on the left of Figure~\ref{fig:nonAsserting}, 
where current assignment for $a,b,c,v$ is respectively $0?,0?,?1,00$. Then we 
decide that $d = 11$, we get (omitting impact mask and decide depth): 
from 3 that $a=00$ with the propagation clause $\bneg a \mid v \mid \bneg d$ ; 
from 2, that   $b=01$  with $b \mid \bneg{d \gg 1}$ ;  from 1, we get $a=01$ 
with $a \mid \bneg c \mid \bneg b$: conflict with mask 01.
Then resolution gives us the clausal constraint $\bneg c \mid v \mid \bneg d \mid 
\bneg{d \gg 1}$, the conclusion of the derivation on the right of 
Figure~\ref{fig:nonAsserting}. After backtracking and applying the values for 
$c$ and $v$, we get $?0 \mid \bneg d \mid \bneg{d \gg 1}$ which does not 
propagate. Indeed, if we look at the level of bits, this gives the clauses 
$\bneg{c_1} \vee \bneg{d_1} $ and $\bneg{d_0} \vee \bneg{d_1} $.

The above clause provokes a problem 
because it  relates \textit{different} positions in the \textit{same} 
variable (the variable $d$ above appears both as a literal with and without 
shift). This sort of problem arises because a decision is allowed to be made on 
many bits of the same variable, unlike in bit-blasting where decisions are made 
on Boolean variables.

To account for these problematic cases, the constraint solver must include in 
the \bv domain representation some indication that the incriminated variable 
(the one which bits are related by the clause, here $d$) can only be subject to 
partial 
decisions. This extra information can be a simple flag that, if set, forces a 
naive bit-by-bit decision, but a more sophisticated information can be 
generated using the conflict mask to derive decision masks associated to the
variable, whereby grouped decisions are allowed on the bits indicated by  that 
mask for that variable. {\it This way, the clause becomes asserting}, in the sense 
that the domain representation of the variable has been propagated in a 
way that prevents the conflict from arising again. 

\subsection{Extended explanations for learning from simplifications}
\label{sec:simplif}
The explanations presented so far fit well bitvector domain propagation, yet 
the solving process  can be greatly improved by 
simplifications and factorizations \cite{chihani17cpaior}. We  explain how the learning presented in 
this paper can be naturally  extended to these extra propagations.

\paragraph{Extended explanations}
 We use $C \genExp E$ to mean that the constraint $C$ can generate the 
explanation containing the propagation clause $E$ and omit decide depth. We can 
now explain using regular Boolean clauses as well, where atoms are formed 
through \bv constraints, equalities or disequalities. Some 
selected simplifications and factorizations are as follows 
(we consider the equality constraint $a=0$ and the \bv literal $\bneg a$ to be 
equivalent):

\begin{itemize}
 \item \vspace{-2mm}
A logical shift constraint $a=a\gg c$ implies the nullity of $a$ if $c>0$. We 
allow each logical shift constraint to generate an explanation for such an 
update: \vspace{-2mm}
\[
 a_1=a_2\gg c \genExp (a_1=a_2)\wedge (c>0) \rightarrow (a_1=a_2=0)
\]
\item \vspace{-2mm}
A disjunction constraint  $a=b_1 \mid b_2$ becomes an equality between $a$ and $b_1$ when $b_2$ becomes 0, and {\it vice versa}, with explanations: \vspace{-2mm}
\[
 a=b_1 \mid b_2 \genExp  b_i=0 \rightarrow (a=b_j) \textit{ with } 
\permof{i,j}{1,2} 
\]
\item \vspace{-2mm}
The exclusive disjunction constraint $a=b\xor c$ is equivalent to any 
permutation of its variables, and   two $\xor$ constraints 
with 2 common variables  prompt an equality between their third respective 
variables, with the following explanation -- where 
$x,y,z$ (resp.~$x',y',z'$) is  a permutation of $a,b,c$ (resp.~$a',b',c'$): \vspace{-2mm}
\[
 a=b\xor c \wedge a'=b'\xor c'  \genExp (x=x') \wedge (y=y') \rightarrow (z=z') 
 \]
\end{itemize}

\vspace{-2mm}
The basic idea is then to combine our  word-level resolution together with regular Boolean resolution in order to derive 
learnt clauses.

\paragraph{Example} We now give an example where, to keep it simple, the explanations cover all the 
conflict mask, such that only one clause is learnt in the end.

\vspace{-2mm}
\begin{align*}
  &(1.) x=x\gg y & \quad  & (2.)z=x\mid t & \quad   & (3.)z=a\xor b  &\quad   &  
(4.) b=c\xor t &\\
    & (5.)a=c \gg y &  &(6.)a=v \band m & &(7.) \bneg a = v \mid n    &&&&
\end{align*}
\vspace{-2mm}

\begin{figure}[htbp]
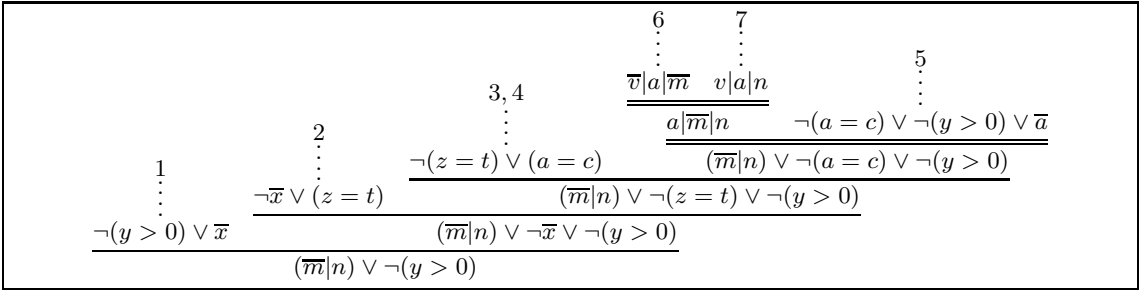
 
\centering
\footnotesize
$\infer{(\bneg m| n) \vee \neg (y>0)}
{\infer*{ \neg (y>0) \vee \bneg x}{1}
&
\infer{(\bneg m| n) \vee \neg  \bneg x\vee \neg (y>0)}{
\infer*{\neg \bneg x \vee  (z=t) }{2}
&
\infer{(\bneg m| n) \vee \neg (z=t) \vee \neg (y>0)}{
\infer*{ \neg (z=t) \vee  (a=c)  }{3,4}
&
\infer={(\bneg m| n) \vee \neg (a=c) \vee \neg (y>0)}
{\infer={a | \bneg m| n}
{\infer*{\bneg v | a | \bneg m}
{6}
&
\infer*{v | a| n}
{7}}
&
\infer*{\neg (a=c) \vee \neg (y>0) \vee \bneg a}{5}
}}}}
$
 \caption{Learning through a combination of \bv- and Boolean- resolution}
 \label{fig:simplification-learning}
\end{figure}

We suppose current known values are $m=1111$ and $n=0000$, and we decide 
on $y=0011$, (or 3, in decimal notation). Then, propagation steps are:
 $x=0$ from 1 because $y=3>0$,
 $z=t$ from 2 because $x$ turned into the neutral element,
 then from 3 and 4 get $a = c$,
 and from that and $y=3>0$ get $a = 0$,
 finally, through  \bv propagation $a=v \band m$ propagates that 
$v=0$
 then $\bneg a = v \mid n$ attempts to change 
that to  $v=1111$: conflict. 
The resolution derivation that allows learning is shown in Figure 
\ref{fig:simplification-learning} (the numbers in the leaves are those of the 
constraints that prompted the respective explanation clause). This 
derivation uses both \bv word-level resolution (marked with a double 
horizontal line) and regular Boolean resolution (marked with a single 
horizontal line). 
The conclusion of that derivation, \ie the learnt clause $(\bneg m| n) \vee \neg 
(y>0) $ (combining both \bv and arithmetic atoms) excludes any value 
strictly greater than 0 from the 
concretization of $y$ in the (current) event 
that $n=0$ and $m=1111$. 

\makeatletter{}\section{Related work} \label{sec:related}

We have already discussed at length the general CDCL approach 
\cite{DBLP:conf/dac/MoskewiczMZZM01,biere2009conflict}  
as well as its straightforward instantiation to bit-level learning over bitvectors 
in Sections \ref{sec:intro} and \ref{sec:background}.  
We present now a few other learning methods from Constraint Programming (CP) -- that can be in some ways adapted to world-level bitvector learning , as well as 
the very few learning methods dedicated to bitvectors.

\paragraph{Learning in Constraint Programming}  There is a number of published work exploring learning capabilities in the 
context of a CP  solver. {\it Lazy Clause Generation}~\cite{ohrimenko07cp} converts 
constraint propagators into clauses and feeds them to a SAT solver, showing 
significant results for scheduling problems. We are not aware of any treatment 
for \bvs in this context but suspect that any such treatment would resort to 
bit-blasting (necessary to use a SAT solver in the background), which is 
precisely what we are trying to avoid. 
{\it (Generalized) NoGood Learning} 
techniques~\cite{dechter1990enhancement,katsirelos05aaai}
improve backtracking search in the context of Constraint Satisfaction Problems. Basically,  (Generalized) NoGood learning 
allows to add partial assignments and non-assignments that are known to 
lead to conflict. This method extends naturally  to word-level learning for  \bvs, 
yet our own approach is more general, as our learnt constraints do not 
mention particular assignments, they are  solely derived  from the 
relationship between the propagation clauses of the constraints that  
lead to a conflict. 
The HaifaCSP solver~\cite{veksler16jai}  is able to learn general constraints  
 through a resolution-inspired method in a  
more general and concise way  
than nogood assignments. The approach uses a  non-clausal 
learning that relies on direct inference between the constraints themselves as 
well as SAT inspired techniques, notably a similar heuristic to 
VSIDS~\cite{Moskewicz01chaff} for variable ordering. It currently handles many 
of the constraints appearing in modern CP solvers, yet bitvector constraints are not handled. 
Being resolution-based able to  
handle integer constraints, this technique is  a good candidate to test the limits 
of the ideas presented in this  paper for theory learning cooperation.

\paragraph{Learning over bitvectors} 
There are two most closely related papers, mentioned in the introduction. 
First, Wang {\it et al.} describes a  
solver~\cite{wang16cpaior} that uses the same word-level propagations as we do 
but delegates  bit-level learning  to a SAT solver. Focusing on the learning 
scheme, which is the subject of our paper, the comparison between our work and 
this one is the same as with any bit-blasting learning approach.  

The closest related work is a model constructing SAT 
solver~\cite{de2013model} whose learning 
mechanism was extended~\cite{zeljic2016deciding} to the theory of quantified 
free bit-vectors, named mcBV and avoiding bit-blasting. The learning mechanism 
of mcBV resembles nogood learning in that it starts from a particular failing 
assignment. The difference is in mcBV's generalization method that starts 
from that assignment and weakens it as much as possible through successive 
flipping of bits from known to unknown. In contrast to our learning, which 
directly involves interactions between propagators, this learning is still 
tied to a particular assignment. Finally, the mcBV approach's handling of 
bounded arithmetic makes it a good candidate for  combination with 
our approach. 

One can also find a relation to the lazy layered~\cite{bruttomesso2007lazy} 
treatment of \qfbv, where bit-blasting is only invoked if necessary (the bits 
of a \bv are all considered equal until those bits are affected differently -- 
through a propagation or through a conflict -- at which moment that 
vector is only blasted as much as necessary). Learning in that setting depends 
on where bit-blasting is done and is still, in essence, Boolean CDCL, even if 
each Boolean variable can potentially represent all bits of a variable (since 
those bits must be equal).

\section{Conclusion} \label{sec:conclusion}

This paper presents a word-level CDCL-style learning mechanism for the theory 
of 
bitvectors, which is of paramount importance in software verification. The 
technique can be integrated in any world-level bitvector solver, whatever the 
underlying technology (CP \cite{chihani17cpaior} or natural-domain SMT 
\cite{cotton2010natural}), as long as the deduction mechanism produces the 
required domain-based explanations and conflict detection.  This work 
illustrates the effect of cross-fertilization between the CP and SMT community 
that brought forth many fruitful advances, most notably CDCL itself. 

The immediate future work is, of course, the implementation of our ideas in a 
word-level solver for practical evaluation and    the fine-tuning of all 
practical ingredients  of an efficient  CDCL-style learning mechanism,  
including  clause deletion, variable ordering, and other learning-related 
heuristics.

\bibliography{myBib}


\end{document}